\input harvmac.tex
\vskip 1.5in
\Title{\vbox{\baselineskip12pt
\hbox to \hsize{\hfill}
\hbox to \hsize{\hfill}}}
{\vbox{
	\centerline{\hbox{(Non)triviality of Pure Spinors
		}}\vskip 5pt
        \centerline{\hbox{and Exact Pure Spinor - RNS Map
		}} } }

\centerline{Dimitri Polyakov\footnote{$^\dagger$}
{twistorstring@gmail.com,dp02@aub.edu.lb; the address after January 5,2009:
National Institute for Theoretical Physics, Department of Physics and
 Centre for Theoretical Physics, University of the Witwatersrand, Wits
2050, South Africa}}
\medskip
\centerline{\it Center for Advanced Mathematical Sciences}
\centerline{\it  American University of Beirut}
\centerline{\it Beirut, Lebanon}
\vskip .3in
\centerline {\bf Abstract}
 All the BRST-invariant operators in pure spinor formalism in $d=10$
can be
represented as BRST commutators, such as
$V=\lbrace{Q_{brst}},{{\theta_{+}}\over{\lambda_{+}}}V\rbrace$
where $\lambda_{+}$ is the $U(5)$ component of the pure spinor
transforming  as  $1_{5\over2}$. Therefore, in order to secure
non-triviality of BRST cohomology in pure spinor string theory,
one has to introduce  ``small Hilbert space'' and
``small operator algebra'' for  pure spinors,
 analogous to those existing in RNS formalism.
As any invariant vertex operator in RNS string theory can also represented
as a commutator $V=\lbrace{Q_{brst}},LV\rbrace$ where 
$L=-4c\partial\xi{\xi}e^{-2\phi}$, we show that
mapping ${{\theta_{+}}\over{\lambda_{+}}}$ to L leads to identification
of the pure spinor variable $\lambda^{\alpha}$ in terms of RNS variables
without any additional non-minimal fields.
We construct the RNS operator satisfying all the properties of 
$\lambda^\alpha$ and show that the pure spinor BRST operator
$\oint{\lambda^\alpha{d_\alpha}}$ is mapped (up to similarity transformation)
to the BRST operator of RNS theory under such a construction.

 {\bf }
{\bf}. 
\Date{October 2008}
\vfill\eject
\lref\berk{N. Berkovits, JHEP 0801:065(2008)}
\lref\nberk{N. Berkovits, JHEP 0004:018 (2000)}
\lref\nberkk{ N. Berkovits, Phys. Lett. B457 (1999)}
\lref\nberkkk{N.Berkovits, JHEP 0108:026 (2001)}
\lref\green{M.B. Green, J.H. Schwarz, Phys. Lett. B136 (1984) 367}
\lref\fms{D. Friedan, E. Martinec, S. Shenker, Nucl. Phys. B271 (1986) 93}
\lref\sieg{W. Siegel, Nucl. Phys. B263 (1986) 93}
\lref\howe{ P. Howe, Phys. Lett. B258 (1991) 141}
\lref\phowe{ P. Howe, Phys. Lett. B273 (1991) 90}
\lref\verl{H. Verlinde, Phys.Lett. B192:95(1987)}
\lref\bars{I. Bars, Phys. Rev. D59:045019(1999)}
\lref\barss{I. Bars, C. Deliduman, D. Minic, Phys.Rev.D59:125004(1999)}
\lref\barsss{I. Bars, C. Deliduman, D. Minic, Phys.Lett.B457:275-284(1999)}
\lref\lian{B. Lian, G. Zuckerman, Phys.Lett. B254 (1991) 417}
\lref\pol{I. Klebanov, A. M. Polyakov, Mod.Phys.Lett.A6:3273-3281}
\lref\wit{E. Witten, Nucl.Phys.B373:187-213  (1992)}
\lref\grig{M. Grigorescu, math-ph/0007033, Stud. Cercetari Fiz.36:3 (1984)}
\lref\witten{E. Witten,hep-th/0312171, Commun. Math. Phys.252:189  (2004)}
\lref\wb{N. Berkovits, E. Witten, hep-th/0406051, JHEP 0408:009 (2004)}
\lref\zam{A. Zamolodchikov and Al. Zamolodchikov,
Nucl.Phys.B477, 577 (1996)}
\lref\mars{J. Marsden, A. Weinstein, Physica 7D (1983) 305-323}
\lref\arnold{V. I. Arnold,''Geometrie Differentielle des Groupes de Lie'',
Ann. Inst. Fourier, Grenoble 16, 1 (1966),319-361}
\lref\self{D. Polyakov,  Int.J.Mod. Phys A20: 2603-2624 (2005)}
\lref\selff{D. Polyakov, Phys. Rev. D65: 084041 (2002)}
\lref\ampf{S.Gubser,I.Klebanov, A.M.Polyakov,
{ Phys.Lett.B428:105-114}}
\lref\malda{J.Maldacena, Adv.Theor.Math.Phys.2 (1998)
231-252, hep-th/9711200} 
\lref\sellf{D. Polyakov, Int. J. Mod. Phys A20:4001-4020 (2005)}
\lref\selfian{I.I. Kogan, D. Polyakov, Int.J.Mod.PhysA18:1827(2003)}
\lref\doug{M. Douglas et.al. , hep-th/0307195}
\lref\dorn{H. Dorn, H. J. Otto, Nucl. Phys. B429,375 (1994)}
\lref\prakash{J. S. Prakash, 
H. S. Sharatchandra, J.Math.Phys.37:6530-6569 (1996)}
\lref\dress{I. R. Klebanov, I. I. Kogan, A. M.Polyakov,
Phys. Rev. Lett.71:3243-3246 (1993)}
\lref\selfdisc{ D. Polyakov, hep-th/0602209, to appear
in IJMPA}
\lref\wittwist{E. Witten, Comm. Math.Phys.252:189-258 (2004)}
\lref\wittberk{ N. Berkovits, E. Witten, JHEP 0408:009 (2004)}
\lref\cachazo{ F. Cachazo, P. Svrcek, E. Witten, JHEP 0410:074 (2004)}
\lref\barstwist{ I. Bars, M. Picon, Phys.Rev.D73:064033 (2006)}
\lref\barstwistt{ I.Bars, Phys. Rev. D70:104022 (2004)}
\lref\selftwist{ D. Polyakov, Phys.Lett.B611:173 (2005)}
\lref\klebwitt{I. Klebanov, E.Witten, Nucl.Phys.B 556 (1999) 89}
\lref\klebgauge{C. Herzog, I. Klebanov, P. Ouyang, hep-th/0205100}
\lref\ampconf{A. M. Polyakov, Nucl.Phys.B486(1997) 23-33}
\lref\amplib{A. M. Polyakov, hep-th/0407209,in 't Hooft, G. (ed.):
50 years of Yang-Mills theory 311-329}
\lref\wit{E. Witten{ Adv.Theor.Math.Phys.2:253-291,1998}}
\lref\alfa{D. Polyakov, Int.J.Mod.Phys.A22:5301-5323(2007)}
\lref\ghost{D. Polyakov, Int.J.Mod.Phys.A22:2441(2007)}
\lref\oda{K. Oda, M. Tonin, Nucl. Phys. B779:63-100(2007)}

\centerline{\bf Introduction}

Pure spinor formalism for superstrings
has been proposed by Berkovits several years  ago 
~{\nberk} as an alternative 
method of covariant quantization of Green-Schwarz superstring theory
~{\green}.
It involves the remarkably simple worldsheet action:

\eqn\grav{\eqalign{
S=\int{d^2z}\lbrace{1\over2}\partial{X_m}\bar\partial{X^m}
+p_\alpha\bar\partial\theta^{\alpha}+{\bar{p_\alpha}}\partial\bar\theta^\alpha
+\lambda_\alpha\bar\partial{w^\alpha}+\bar\lambda_\alpha
\partial{\bar{w^\alpha}}\rbrace}}
where $p_\alpha$ is conjugate to $\theta_\alpha$ ~{\sieg}
and the commuting spinors $\lambda^\alpha$
and $w^\alpha$ are the bosonic ghosts which, roughly speaking,
are related to the fermionic gauge $\kappa$-symmetry in GS superstring
theory. The action (1) is related to  the standard GS action
by substituting the constraint

\eqn\lowen{d_\alpha=p_\alpha-{1\over2}(\partial{X^m}+
{1\over4}\theta\gamma^m\partial\theta)(\gamma^m\theta)_\alpha=0}

and the corresponding BRST operator
\eqn\lowen{Q_{brst}=\oint{{dz}\over{2i\pi}}\lambda^\alpha{d_\alpha}(z)}
is nilpotent provided that $\lambda^\alpha$ satisfies the pure spinor 
condition:

\eqn\lowen{\lambda^\alpha\gamma^m_{\alpha\beta}\lambda^\beta=0}
reducing the number of independent components of $\lambda$
from 16 to 11.
An example of unintegrated massless vertex operator 
in such a BRST cohomology is given by

\eqn\lowen{U=\lambda^\alpha{A_\alpha}(X,\theta)}.

This operator is physical provided that the space-time superfield
$A_\alpha$ is on-shell:

\eqn\lowen{
\gamma_{m_1...m_5}^{\alpha\beta}D_\alpha{A}_\beta=0}

(this particularly implies the standard Maxwell equation for the bosonic 
vector component of $A$)
and thus the vertex operator (5) is identified with the emission
of a photon by the superstring ~{\nberk, \howe, \phowe}.
The integrated version of this operator $\sim\oint{{dz}\over{2i\pi}}{V}(z)$
satisfying $\lbrack{Q_{brst}},V\rbrack=\partial{U}$ can also be constructed,
with $V$ obviously having ghost number zero ~ {\nberkkk}.
Physical vertex operators (both massless and massive)
considered in pure spinor formalism thus typically have ghost number 1 in 
unintegrated form and number zero in the integrated version.

The important question is how the PS approach is related to
other descriptions of superstring , such as RNS formalism. 
While such a relation exists
and can be constructed, the construction is
 not straightforward and the constructions considered so far particularly
required the introduction of additional non-minimal fields
by hands ~{\berk, \nberkkk}

 Another natural question is whether the PS superstring could contain any
additional physical operators, e.g. with higher ghost numbers.
It is far from obvious that such operators could exist at all.
For example, a straightforward naive attempt to  generalize the 
unintegrated operator (5) to the ghost number $2$ case:

\eqn\lowen{U_2=\lambda^{\alpha}\lambda^{\beta}
{F_{\alpha\beta}}(X,\theta)}

fails as the on-shell conditions for the field $F_{\alpha\beta}$:

\eqn\lowen{\gamma_{m_1...m_5}^{\alpha\beta}D_\gamma{F_{\alpha\beta}}(X,\theta)
=0}

imply the triviality of the $U_2$ operator:

\eqn\lowen{U_2=\lbrace{Q_{brst}},\theta^\alpha\lambda^\beta{F_{\alpha\beta}}
\rbrace}
Similarly, naive construction of  ghost number $n$ operators 
$\sim\lambda^n$ leads to BRST-exact expressions, provided the on-shell
constraint on the corresponding background fields.
Despite that, below we shall demonstrate that vertex operators
with non-standard coupling to pure spinors do appear in BRST cohomology.
In general,
 the question of non-triviality of BRST cohomology 
in the PS formalism appears more subtle compared to RNS.
That is, since $\lbrace{Q},\theta^\alpha\rbrace=\lambda^\alpha$
and  $\lbrack{Q},\lambda^\alpha\rbrack=0$, any invariant operator $V$
in pure spinor string theory can be  written as an exact BRST commutator.
For example, consider the standard U(5)-invariant parametrization of 
$\lambda^\alpha$: $\lambda^\alpha=(\lambda^{+},\lambda^{ab},\lambda^a)
(a,b=1,..,5)$ with 
$\lambda_{ab}=-\lambda_{ba}$ and
$\lambda^a=\epsilon^{abcde}\lambda_{bc}\lambda_{de}$.
Then any invariant $V$ can be written as

\eqn\lowen{V=\lbrack{Q_{brst}},
{{\theta_{+}}\over{{\lambda^{+}}}}V\rbrack}.

This poses a question whether 
BRST cohomology of PS string theory is empty
(similar observations have also been made in ~{\oda})

In fact, the identity (10) is reminiscent of the similar relation
in the RNS formalism where any invariant $V$ can be written as
\eqn\lowen{V=\lbrace{Q_{brst}^{RNS}},LV\rbrace}
where
\eqn\lowen{L=-4ce^{2\chi-2\phi}=-4c\partial\xi\xi{e^{-2\phi}}}
with the ghost fields bosonized as ~{\fms}
\eqn\brav{\eqalign{b=e^{-\sigma},c=e^{\sigma}\cr
\beta=e^{\chi-\phi}\partial\chi\equiv\partial\xi{e^{-\phi}},
\gamma=e^{\phi-\chi}}}
It is easy to check
\eqn\lowen{\lbrace{Q_{brst}^{RNS}},L\rbrace=1}
In RNS approach, however, the relation (12) does not lead to 
the triviality of states since the $L$-operator is not in
 the small operator algebra, as it
explicitly depends on $\xi=e^\chi$ (rather than its derivatives).
So the only way to bail out pure spinors 
is to introduce similar classification
for the PS formalism as well.
Such a  classification, however, isn't as obvious as in the RNS case.
In the RNS case we exclude the operators with explicit
$\xi$-dependence because the bosonization relations
for the ghost fields $\beta$ and $\gamma$ depend on the derivative
of $\xi$, but not on $\xi$ itself ($\xi$ can only be expressed as a
 generalized step function of $\beta$: $\xi=\Theta(\beta)$)
In the PS formalism, however, the analogue of the $L$-operator
is given by the ratio
${{\theta_{+}}\over{{\lambda^{+}}}}$ consisting of fields 
already present in the theory. For this reason, the 
distinction between ``large'' and ``small'' operator algebras appears
more obscure in the PS approach.
One possible approach is to try to construct a direct map between
$PS$ and $RNS$ variables, which in particular would identify 
${{\theta_{+}}\over{{\lambda^{+}}}}$ with the $L$-operator
of RNS formalism.
Once such a map is constructed, it would transform
the states from the little Hilbert space in
RNS formalism to those in the little Hilbert space in the pure spinor
 description.
So we start with the map
\eqn\lowen{c\partial\xi\xi{e^{-2\phi}}\sim{{\theta_{+}}\over{{\lambda^{+}}}}}
and will try to deduce the correspondence between 
PS ans RNS variables by using this isomorphism.
Since the Green-Schwarz variable $\theta^\alpha$ is known to be
related to RNS spin operator by the field redefinition 
\eqn\lowen{\theta^\alpha
\sim{e^{\phi\over2}}\Sigma^\alpha}, we write
$\theta^{+}={e^{\phi\over2}}\Sigma^{+}$
where $\Sigma^{+}$ is the component of $\Sigma^\alpha$ with
five pluses $(+++++)$  in the $(\pm)^5$ representation.
For our purposes, it is convenient to split 32-component 
spin operator into two 16-component spin operators $\Sigma_\alpha$
and ${\tilde\Sigma}_\alpha$ with opposite GSO parities.
Then the RNS expression for 
${{1}\over{{\lambda^{+}}}}\equiv(\lambda^{+})^{-1}$ 
which OPE with $\theta^{+}$ gives $L$
is given by
\eqn\lowen{(\lambda^{+})^{-1}=ce^{2\chi-{5\over2}\phi}{\tilde\Sigma}^{+}}
where 
${\tilde\Sigma}^{+}$ is the $(-----)$ component of the
32-component spin operator (so it has GSO parity opposite to $\Sigma^{+}$).
One can easily check that the OPE of 
$(\lambda^{+})^{-1}$ with $\theta^{+}$
is non-singular, with the zero order term given precisely by $L$.
Next, the $\lambda^{+}$ operator can be read off the OPE
\eqn\lowen{(\lambda^{+})^{-1}(z)\lambda^{+}(w)\sim
1+O(z-w)}
It is easy to see that
\eqn\lowen{\lambda^{+}=b{e}^{{5\over2}\phi-2\chi}\Sigma^{+}}
is precisely the operator satisfying this OPE identity.
Note that $\lambda^{+}$ and $\theta^{+}$ have the same GSO parity.
It is now straightforward to identify

\eqn\lowen{\lambda^\alpha{\sim}b{e}^{{5\over2}\phi-2\chi}\Sigma^{\alpha}}
however such an identification is not yet complete for the following reason.
On one hand, the expression (20) of $\lambda^\alpha$ in terms
of RNS variables does have 
some basic properties of pure spinors:
it is the dimension zero primary field, it is a commuting spinor
(since it is multiplied by the b-ghost which is worldsheet fermion)
however its full OPE does not yet satisfy the pure spinor constraint as
\eqn\grav{\eqalign{\lambda^\alpha(z)\lambda^\beta(w)\sim
{1\over{(z-w)^2}}\partial{b}be^{5\phi-4\chi}\gamma^m_{\alpha\beta}\psi_m
+{1\over4}\partial{b}be^{5\phi-4\chi}\gamma^m_{\alpha\beta}
\partial^2\psi_m\cr
+\partial{b}be^{5\phi-4\chi}\gamma_{m_1...m_5}^{\alpha\beta}\psi^{m_1}
...\psi^{m_5}+...}}
so the while the second term of the normally ordered part of this OPE
would vanish after substituting into the left hand side 
of the pure spinor
constraint ~ $\lambda\gamma^m\lambda$
 (since it would produce the factor proportional to
$\sim{Tr(\gamma^m\gamma^{m_1...m_5})=0}$, the first
 term would still contribute.
In addition, the OPE (21) has a double pole singularity while
the OPE of two $\lambda$'s in the pure spinor formalism is known to be
non-singular ~{\nberkkk}
The reason is that both the OPE singularity and the violation
of the pure spinor constraint are related to BRST non-invariance of 
the operator (20), while the
actual pure spinor must be BRST-invariant.
For this reason, one has to add the correction terms to the r.h.s.
of (20) to ensure the BRST-invariance. This can be done by replacing
\eqn\lowen{\lambda^\alpha\rightarrow\lambda^\alpha-L\rho^\alpha}
where $\rho^\alpha=\lbrack{Q_{brst}},
b{e}^{{5\over2}\phi-2\chi}\Sigma^{\alpha}\rbrack$ is the
BRST commutator with the right hand side of (20).
Since $\lbrace{Q_{brst}},L\rbrace=1$ and $\lbrack{Q_{brst}},\rho^\alpha\rbrack
=0$, the modified $\lambda^\alpha$ will be BRST-invariant by construction.
Evaluating $\rho^\alpha$ and its normally ordered product with $L$
we find the complete RNS representation for the pure spinor
variable $\lambda^\alpha$ to be given by
\eqn\grav{\eqalign{
\lambda_\alpha=be^{{5\over2}\phi-2\chi}\Sigma_\alpha
+2e^{{3\over2}\phi-\chi}\gamma^m_{\alpha\beta}\partial{X_m}
{\tilde\Sigma}^\beta
-2ce^{{1\over2}\phi}\Sigma_\alpha\partial\phi
-4ce^{{1\over2}\phi}\partial\Sigma_\alpha}}
It is straightforward to check that this expression for  $\lambda^\alpha$ does
satisfy the pure spinor condition (4) (see the Appendix).
Note that $\lambda^\alpha=-4{\lbrace}Q_{brst},\theta^\alpha\rbrace$
where the factor of $-4$ is related to our normalization choice
in (15).
Note that
$\lambda^\alpha$ is annihilated by inverse picture-changing operator
$\Gamma^{-1}=4c\partial\xi{e^{-2\phi}}$ and therefore cannot
be transformed to pictures lower than ${1\over2}$, such as $-{1\over2}$
or $-{3\over2}$.
In the next section we will use the RNS expression (23) for
$\lambda^\alpha$ in order to map the BRST charge of
pure spinor string theory into RNS BRST charge.

\centerline{\bf RNS BRST Operator from Pure Spinor BRST Operator}

In this section we will use the RNS representation (23) for the pure
spinor variable $\lambda^\alpha$ to construct the exact map
relating pure spinor BRST charge and RNS BRST charge.
To demonstrate this relation we have to calculate the normally ordered
expression of the pure spinor BRST current 
$:\lambda^\alpha{d_\alpha}:$ in the RNS formalism.
The constraint operator 
\eqn\lowen{
d_\alpha=p_\alpha-{1\over2}\theta^\beta\gamma_m^{\alpha\beta}\partial{X_m}
-{1\over8}(\theta\gamma^m\partial\theta)(\gamma_m\theta)_\alpha}
consists of three terms, so we are to calculate
the normally ordered OPE's of these terms with $\lambda^\alpha$
one by one.
A useful formula for our calculation is the OPE between two spin 
operators:
\eqn\grav{\eqalign{\Sigma_\alpha(z)\Sigma_\beta(w)
\sim{{{{\gamma}^m_{\alpha\beta}\psi_m}({{z+w}\over2})}\over{(z-w)^{{3\over4}}}}
+{1\over6}(z-w)^{1\over4}{{\gamma}^{mnp}_{\alpha\beta}\psi_m\psi_n\psi_p}
({{z+w}\over2})+...\cr
\Sigma_\alpha(z){\tilde\Sigma}_\beta(w)
\sim{{\delta_{\alpha\beta}}\over{(z-w)^{5\over4}}}
+{1\over2}(z-w)^{3\over4}\partial\psi_m\psi^m({{z+w}\over2})+...}}
where we skipped higher order OPE terms as well as those
not contributing to the normally ordered expression
for $:\lambda^\alpha{d_\alpha}:$.
Note that, as the ordered RNS expressions for three terms (24) of
$d_\alpha$ contain zero, one and three gamma-matrices
 respectively (see below), only the terms with one or three 
gamma-matrices in the $\Sigma\Sigma$ or ${\tilde\Sigma}{\tilde\Sigma}$
operator products and only
the terms proportional to $\delta_{\alpha\beta}$
in the $\Sigma{\tilde\Sigma}$ OPE contribute
to the BRST current. All other OPE terms
(i.e. those with the number of antisymmetrized gamma-matrices
other than 0,1 or 3) are irrelevant to us since their contributions
to $:\lambda^\alpha{d_\alpha}:$ produce terms proportional to vanishing
 traces of antisymmetrized gamma-matrices.

We start with the $p_\alpha$ term of $d_\alpha$
Since $p_\alpha$ is canonically conjugate to $\theta^\beta$:
\eqn\lowen{p_\alpha(z)\theta^\beta(w)\sim{{\delta_\alpha^\beta}\over{z-w}}}
the RNS representation for $p_\alpha$ is easily deduced to be
\eqn\lowen{p_\alpha={e^{-{1\over2}\phi}}{\tilde\Sigma}_\alpha}
i.e. it is simply the space-time supercurrent at picture ${-{1\over2}}$.
Then the normally ordered
 OPE's of $p_\alpha$ with the first two terms of $\lambda^\alpha$
are easily evaluated to give

\eqn\grav{\eqalign{p_\alpha(z)b{e^{{5\over2}\phi-2\chi}\Sigma^\alpha}(w)
={(z-w)^0}b{e^{2\phi-2\chi}}({{z+w}\over2})+O(z-w)\cr
p_\alpha(z)2e^{{3\over2}\phi-\chi}\gamma^{m}_{\alpha\beta}{\tilde\Sigma}(w)
=(z-w)^0{e^{\phi-\chi}}\psi^m\partial{X_m}({{z+w}\over2})+O(z-w)}}

so the result is given by easily recognizable (up to normalization factors)
ghost and matter supercurrent terms of $j_{brst}$ in the RNS formalism.
The OPE of $p_\alpha$ with the remaining part of $\lambda^\alpha$,
namely,
$ce^{{1\over2}\phi}\Sigma_\alpha(2\partial\sigma-2\partial\phi)
-6ce^{{1\over2}\phi}\partial\Sigma_\alpha$
is a bit more tedious but straightforward to calculate
producing terms with the structure $\sim{c}G_{(2)}(\psi,\sigma,\phi,\chi)$
with $G_2$ being an operator of conformal dimension two, consisting of
$\psi,\phi,\chi$ and $\sigma$ worldsheet fields, giving
a hint on the relevance of this contribution to the 
$cT+b\partial{c}{c}$ part of $Q_{brst}$ in the RNS description.
Performing the calculation and collecting all the terms together
we obtain the contribution of $:p_\alpha\lambda^\alpha:$ to $j_{brst}$
to be 
given by
\eqn\grav{\eqalign{:p_\alpha\lambda^\alpha:
=\delta_\alpha^\alpha\lbrace{\gamma^2b}+\gamma\psi^m\partial{X_m}
\cr
+c({{11\over4}}\partial\psi^m\psi_m
-{13\over{16}}(\partial\phi)^2
+\partial^2\phi+{1\over{16}}(\partial\sigma)^2-{{15}\over{16}}
\partial^2\sigma-{3\over4}\partial\phi\partial\sigma)\rbrace}}

The next step is to calculate the contribution stemming from the normally
ordered OPE of $\lambda^\alpha$ with  the second term of $d_\alpha$, given by
$-{1\over2}\partial{X_m}(\gamma^m\theta)_\alpha$. However, an
important remark should be made first. Since the RNS expressions
 (16) for
$\theta_\alpha$ and (23) for 
$\lambda^\alpha$
are both at the ghost picture ${1\over2}$, the straightforward
evaluation of their OPE would give an operator at picture 1.
This is not quite what we are looking for 
since all the terms of $j_{brst}$ are at picture zero.
Since we expect that the OPE of $\lambda^\alpha$
and $-{1\over2}\partial{X_m}(\gamma^m\theta)_\alpha$
reproduces only a part of $j_{brst}$, the resulting operator
is generally off-shell, so one cannot picture transform it in a
 straightforward manner. As for $\lambda^\alpha$,
although it is on-shell, inverse picture-changing
still isn't applicable to it,
as was noted above. For this reason, in order to get a picture zero result
for this contribution,
instead of taking $\theta^\alpha$  in the  standard form (16)
one has to take it in its equivalent form
\eqn\lowen{\theta_\alpha=-4{c}e^{\chi-{3\over2}\phi}\Sigma_\alpha}
which is at picture $-{1\over2}$.
Although picture-changing transformation isn't well-defined for off-shell
variables such as $\theta^\alpha$, the expressions  (16) and (30)
are equivalent since they both satisfy the same canonical relation
with the conjugate momentum $p_\alpha$.
Indeed, since the worldsheet integral of $p_\alpha$ is
on-shell, one can transform it
to picture ${1\over2}$ obtaining
\eqn\lowen{p_\alpha=-{1\over2}e^{{1\over2}\phi}\gamma^m_{\alpha\beta}
{\Sigma}^\beta\partial{X_m}
-{1\over4}be^{{3\over2}\phi-\chi}{\tilde\Sigma}_\alpha}.
Applying $p_\alpha$ of (31) to $\theta^\beta$ of (30) one easily finds
that, while the first term of $p_\alpha$ doesn't contribute
to the simple pole of its OPE with $\theta^\beta$, the second
term's OPE with $\theta$ produces precisely the simple pole leading
to the standard canonical relation.
Thus 
\eqn\lowen{
-{1\over2}\theta^\beta\gamma^m_{\alpha\beta}\partial{X_m}
=2ce^{\chi-{3\over2}\phi}\Sigma^\beta\gamma^m_{\alpha\beta}\partial{X_m}}

Evaluating the OPE of this term with $\lambda^\alpha$ of (23) we
obtain

\eqn\grav{\eqalign{
2ce^{\chi-{3\over2}\phi}\Sigma^\beta\gamma^m_{\alpha\beta}\partial{X_m}
(z)be^{{5\over2}\phi-2\chi}\Sigma^\alpha(w)
=(z-w)^0{\delta_\alpha^\alpha}e^{\phi-\chi}\psi^m\partial{X_m}
({{z+w}\over2})+O(z-w)}}
for the product of (32) with the first term of $\lambda^\alpha$
\eqn\grav{\eqalign{
2ce^{\chi-{3\over2}\phi}\Sigma^\beta\gamma^m_{\alpha\beta}\partial{X_m}(z)
2e^{{3\over2}\phi-\chi}{\tilde\Sigma}^\gamma\gamma^m_{\alpha\gamma}
\partial{X_m}
(w)\cr
=
\delta_\alpha^\alpha{c}{\lbrace}2\partial{X_m}\partial{X^m}
-8\partial\psi_m\psi^m
-2\partial^2\sigma-2(\partial\sigma)^2\cr
-18(\partial\phi)^2
-8(\partial\chi)^2+24\partial\chi\partial\phi-8\partial\chi\partial\sigma
+18\partial\phi\partial\sigma\rbrace({{z+w}\over2})+O(z-w)}}
for the product of (32) with the second term of $\lambda^\alpha$
\eqn\grav{\eqalign{
2ce^{\chi-{3\over2}\phi}\Sigma^\beta\gamma^m_{\alpha\beta}\partial{X_m}
(z)
-2ce^{{1\over2}\phi}\Sigma_\alpha\partial\phi
-4ce^{{1\over2}\phi}\partial\Sigma_\alpha)(w)
\cr
=-\delta_\alpha^\alpha
{7\over2}\partial{c}ce^{\chi-\phi}\psi_m\partial{X^m}({{z+w}\over2})
+O(z-w)
}}
for the product of (32) with the third term of $\lambda^\alpha$

Note the appearance of an extra $\gamma\psi_m\partial{X^m}$ term
on the r.h.s. of the OPE (33) that ensures the correct normalisation
of the matter supercurrent term with respect to the ghost supercurrent term
in $j_{brst}$.
The final contribution to $j_{brst}$ comes from the OPE of
$\lambda^\alpha$ and $-{1\over8}(\theta\gamma^m\partial\theta)
(\gamma_m\theta)_\alpha={1\over8}\theta^\beta\theta^\rho
\partial\theta^\lambda\gamma^{m}_{\beta\lambda}(\gamma_m)_{\alpha\rho}$.
To ensure that the contribution of this OPE to $j_{brst}$ is at picture zero,
it is convenient to take $\theta_\beta$ and $\theta_\rho$
at the picture $-{1\over2}$ representation (30) while keeping
$\partial\theta_\lambda$ at the picture ${1\over2}$ version (16).
Using the OPE (25) one easily finds
\eqn\lowen{:\theta_\beta\theta_\rho:(z)={8\over3}\gamma^{mnp}_{\beta\rho}
\partial{c}c{e^{2\chi-3\phi}}\psi_m\psi_n\psi_p(z)}
Calculating the operator product of (36) with 
$\partial\theta^\lambda=\partial(e^{{1\over2}\phi}\Sigma^\lambda)$
using (25) gives
\eqn\lowen{
-{1\over8}:(\theta\gamma^m\partial\theta)
(\gamma_m\theta)_\alpha:(z)
=-32\partial{c}ce^{2\phi-{5\over2}\phi}(\partial\Sigma_\alpha-{{19}\over6}
\partial\phi\Sigma_\alpha)(z)}
The calculation of the OPE of (37) with the first term of $\lambda^\alpha$
(23)
gives
\eqn\grav{\eqalign{
-32\partial{c}ce^{2\phi-{5\over2}\phi}(\partial\Sigma_\alpha-{{19}\over6}
\partial\phi\Sigma_\alpha)(z)be^{5\phi-2\chi}\Sigma^\alpha(w)\cr
=(z-w)^0\delta_\alpha^\alpha
c{\lbrace}{29\over4}\partial\psi_m\psi^m-{{179}\over{16}}(\partial\phi)^2
-13\partial^2\phi+10(\partial\chi)^2+14\partial^2\chi
\cr
-{{169}\over{16}}
(\partial\sigma)^2+{{167}\over{16}}\partial^2\sigma
+24\partial\chi\partial\phi+{{59}\over4}\partial\phi\partial\sigma
-16\partial\chi\partial\sigma\rbrace({{z+w}\over2})+O(z-w)}}
The OPE of (37) with the second term of $\lambda^\alpha$ gives
\eqn\lowen{-32\partial{c}ce^{2\phi-{5\over2}\phi}
(\partial\Sigma_\alpha-{{19}\over6}
\partial\phi\Sigma_\alpha)(z)2e^{{3\over2}\phi-\chi}\gamma_m^{\alpha\beta}
\partial{X^m}{\tilde\Sigma}^\beta=-{{25}\over2}\partial{c}c{e^{\chi-\phi}}
\psi_m\partial{X^m}(z)+O(z-w)}
Finally, the OPE of (37) with the third term of $\lambda^\alpha$ produces
\eqn\grav{\eqalign{
-32\partial{c}ce^{2\phi-{5\over2}\phi}(\partial\Sigma_\alpha-{{19}\over6}
\partial\phi\Sigma_\alpha)(z)ce^{{1\over2}\phi}(-4\partial\Sigma^{\alpha}
-2\Sigma^\alpha\partial\phi)(w)\cr
=32\partial^2{c}\partial{c}{c}e^{2\chi-2\phi}({{z+w}\over2})+O(z-w)}}
Collecting together all the terms in (29) - (40) we find the overall
normally ordered product of $d_\alpha$ and $\lambda^\alpha$ to be given by:
\eqn\grav{\eqalign{{1\over{16}}:\lambda^\alpha{d}_\alpha:
=2\gamma\psi_m\partial{X}^m+\gamma^2b+c{\lbrace}
2\partial{X_m}\partial{X^m}+2\partial\psi_m\psi^m\cr
-18(\partial\chi)^2
+14\partial^2\chi-30(\partial\phi)^2-12\partial^2\phi
-{{25}\over2}(\partial\sigma)^2+{15\over2}\partial^2\sigma
\cr
+48\partial\chi\partial\phi-24\partial\chi\partial\sigma+32\partial\phi
\partial\sigma\rbrace-16\partial{c}c{e}^{\chi-\phi}\psi_m\partial{X^m}
+32\partial^2{c}\partial{c}c{e^{2\chi-2\phi}}
}}
where the factor of ${1\over{16}}$ in front of the pure spinor
BRST current is to absorb the factor of $\delta_\alpha^\alpha=16$
always appearing on the right hand side of the operator products
(29) - (40). 
Although the RNS expression (41) for the pure spinor
BRST current
looks tedious, it is straightforward to check that,
up to an overall numerical factor and
BRST trivial terms, it is equivalent
to the BRST current in RNS formalism.
Indeed, using the bosonized expression for RNS BRST current:
\eqn\grav{\eqalign{j_{brst}^{RNS}=cT+b\partial{c}c-{1\over2}\gamma
\psi_m\partial{X^m}-{1\over4}b\gamma^2\cr
=c\lbrace-{1\over2}\partial{X_m}\partial{X^m}
-{1\over2}\partial\psi_m\psi^m-{1\over2}(\partial\phi)^2
-\partial^2\phi+{1\over2}(\partial\chi)^2+{1\over2}\partial^2\chi\cr
+{9\over8}(\partial\sigma)^2+{1\over8}\partial^2\sigma\rbrace
-{1\over2}e^{\phi-\chi}\psi_m\partial{X^m}-{1\over4}be^{2\phi-2\chi}}}
and the commutator:
\eqn\grav{\eqalign{
\lbrack{Q_{brst}^{RNS}},\partial{c}c{e^{2\chi-2\phi}\partial\chi}\rbrack
=\partial^2{c}\partial{c}{c}e^{2\chi-2\phi}-{1\over2}\partial{c}c
{e^{\chi-\phi}}\psi_m\partial{X^m}\cr
-{1\over4}c\lbrace
2\partial^2\phi-2\partial^2\chi-\partial^2\sigma
+4(\partial\phi)^2+2(\partial\chi)^2+(\partial\sigma)^2
-6\partial\chi\partial\phi+3\partial\chi\partial\sigma-
4\partial\phi\partial\sigma\rbrace}}
one easly finds 
\eqn\lowen{{1\over{16}}j_{brst}^{pure spinor}=
-4j_{brst}^{RNS}+32
\lbrack{Q_{brst}^{RNS}},\partial{c}c{e^{2\chi-2\phi}}\partial\chi\rbrack}
This concludes the calculation identifying the BRST charges
in RNS and pure spinor approaches. Note that
a shift of a BRST charge by  any BRST trivial term
(that particularly occurs in (44)):

\eqn\lowen{Q_{brst}\rightarrow{Q_{brst}}+\lbrack{Q}_{brst},R\rbrack}
where $R$ is some operator, is equivalent to the similarity
transformation
\eqn\lowen{Q_{brst}\rightarrow{e^{-R}}Q_{brst}e^{R}}
considered in ~{\berk}.
In our case,
\eqn\lowen{R=32\oint{{dz}\over{2i\pi}}
\partial{c}c{e}^{2\chi-2\phi}\partial\chi(z)}

Note that the $R$-operator isn't generally required to be in the ``small 
operator algebra'' and, as a matter of fact, both the $R$-operator
(47) and the $R$-operator used in the similarity transformation in ~{\berk}
are outside the small algebra:
the $R$-operator (47) contains the factor of $e^{2\chi}\partial\chi
={1\over2}\partial^2\xi\xi$, while the R-operator used by Berkovits
explicitly depends on ${1\over{\lambda_{+}}}$ which, when translated
into RNS language, isn't in the small algebra as well.

\centerline{\bf Discussion. Vertex Operators with Non-trivial Pure Spinor
Couplings}

In this letter we have proposed an exact map expressing
the pure spinor variable $\lambda^\alpha$ in terms of 
BRST invariant RNS operator of conformal dimension zero, 
satisfying pure spinor constraint. 
The map is based on identifying the ${{\theta_{+}}\over{\lambda_{+}}}$
operator in the pure spinor formalism
and the $L$-operator in the RNS description
satisfying $\lbrace{Q_{brst}},L\rbrace=1$
This map particularly leads to the identification (23) of 
pure spinor and RNS BRST operators, up to similarity
transformation (or BRST-trivial terms).
The non-triviality of vertex operators in pure spinor approach
requires the introduction of ``small'' and ``large'' operator algebra
in pure spinor approach, similarly to the classification
existing in RNS approach.
However, classifying the full operator algebra
 in terms of ``large'' and ``small'' appears somewhat
ambiguous in the pure spinor formalism,
compared to RNS formalism, where such a classification is clear
and is based on the bosonization relations for superconformal ghosts.
The small operator algebra of the pure spinor formalism
should particularly exclude operators inverse to pure spinor
components, such as ${1\over{\lambda_{+}}}$, but such a constraint
appears too relaxed and also somewhat artificial since, unlike
RNS variable $\xi$, which can only be expressed as a generalized
step function of superconformal $\beta$-ghost: $\xi=\Theta(\beta)$
pure spinor operator ${1\over{\lambda_{+}}}$ is the function
of a variable manifestly present in the theory. This
particularly leads to the pure spinor BRST cohomology
containing operators which physical meaning is unclear.
In particular any function $F(\lambda)$ is an invariant operator
in pure spinor formalism. If $F(\lambda)$ is polynomial,
e.g. $F(\lambda)\sim\lambda_{\alpha_1}...\lambda_{\alpha_n}$, it can be 
represented as a BRST commutator
$F(\lambda)=\lbrace{Q_{brst}},\theta_{\alpha_1}\lambda_{\alpha_2}...
\lambda_{\alpha_n}\rbrace$, i.e. it is BRST exact. If, however, $F(\lambda)$
isn't a polynomial function (e.g. $F(\lambda)\sim{log}(\lambda)$)
then the only way to represent it as a BRST commutator seems to be
$F(\lambda)=\lbrace{Q_{brst}},
{{\theta_{+}}\over{\lambda_{+}}}F(\lambda)\rbrace$,
but this doesn't make an operator unphysical, due to
the small/large algebra classification.
Apparently not all these operators, while formally in the cohomology,
are of physical significance.
For this reason, one needs to find the way 
to eliminate these clearly excessive states, which apparently
requires better understanding of how operator formalism works
in the pure spinor approach. 

\centerline{\bf Acknowledgements}

As I'm preparing to join
the research group at the University of the Witwatersrand
in Johannesburg,
I wish to express my heartfelt gratitude to Wafic Sabra and other members
of Center for Advanced Mathematical Sciences (CAMS) in Beirut
for many illuminating discussions and for the friendship during the years
I have spent in Lebanon.

\centerline{\bf Appendix}

In this short appendix we demonstrate that the BRST-invariant 
RNS expression (23) for $\lambda^\alpha$ satisfies the pure spinor
condition (4). Since $\lambda^\alpha\sim\lbrace{Q_{brst}},\theta^\alpha\rbrace$,
it is sufficient to show
 that the operator $N^m=(\theta\gamma^m\lambda)$
is BRST-invariant.
Taking $\theta^\alpha=e^{{1\over2}\phi}\Sigma^\alpha$ according to (16)
and evaluating its OPE with $\lambda^\alpha$ of (23) using (25)
it is straightforward to calculate

\eqn\grav{\eqalign{-{1\over4}:\theta\gamma^m\lambda:=-{1\over4}
\gamma^m_{\alpha\beta}:e^{{1\over2}\phi}\Sigma^{\alpha}(z)
{\lbrack}be^{{5\over2}\phi-2\chi}\Sigma_\beta
\cr
+2e^{{3\over2}\phi-\chi}\gamma^m_{\beta\lambda}\partial{X_m}
{\tilde\Sigma}^\lambda
-2ce^{{1\over2}\phi}\Sigma_\beta\partial\phi
-4ce^{{1\over2}\phi}\partial\Sigma_\beta{\rbrack}(z):
\cr
=
-{1\over2}e^\phi(\psi_n\partial{X^n})\partial{X^m}
+{1\over2}ce^\phi\lbrack{1\over2}\partial^2\psi^m+\partial\psi^m
(\partial\phi-\partial\chi)+{1\over2}(\partial^2\phi-\partial^2\chi\cr
+(\partial\phi-\partial\chi)^2)\psi^m\rbrack
-{1\over4}\partial^2c{e^\phi}\psi^m+{1\over2}c\partial(e^\phi\partial\chi\psi^m)
\cr
-{1\over8}e^{2\phi-\chi}{\lbrack}2\partial^2\phi+2\partial^2\chi-\partial^2\sigma
+(2\partial\phi-2\partial\chi-\partial\sigma)^2\rbrack\partial{X^m}\cr
-{1\over4}e^{2\phi-\chi}{\lbrack}(\psi_n\partial{X^n})\partial\phi
+\partial(\psi_n\partial{X^n})\psi^m
-\partial\chi\partial^2{X^m}-\partial\chi(\partial\phi-\partial\chi)
\partial{X^m}\cr
+{1\over2}\partial^3{X^m}+(\partial\phi-\partial\chi)\partial^2{X^m}
+{1\over2}(\partial^2\phi-\partial^2\chi+(\partial\phi-\partial\chi)^2)
\partial{X^m}\rbrack\cr
-{1\over8}be^{3\phi-2\chi}\lbrack(2\partial\phi-2\partial\chi-\partial\sigma)
(2\partial\phi-\partial\chi-\sigma)+2\partial^2\phi-2\partial^2\chi
-\partial^2\sigma\rbrack
\psi^m\cr
-{1\over4}\lbrack{Q_{brst}},e^\phi(\psi^m\partial\phi+{1\over2}\partial\psi^m)
\rbrack
}}
Up to the BRST trivial terms, the right hand side of
(48) can be recognized as $\lbrack{Q_{brst}},\xi{V_{photon}}\rbrack$
where $V_{photon}=c\partial{X^m}+{1\over2}\gamma\psi^m$
is the unintegrated photon vertex operator at zero momentum.
For this reason, the RNS expression for
$-{1\over4}\theta\gamma^m\lambda$ is given by the unintegrated
photon vertex operator at superconformal ghost picture 1
at zero momentum (which of course is BRST-invariant)
plus BRST trivial terms.
Therefore $-{1\over4}\theta\gamma^m\lambda$ is BRST-invariant
and its BRST commutator, given by $\lambda\gamma^m\lambda$, is 
identically zero. This concludes the proof that
 $\lambda^\alpha$ satisfies the pure spinor constraint (4).

\vfill\eject

\listrefs
\end